\documentclass[aps, prl, twocolumn, floatfix]{revtex4-2}
\usepackage{graphicx, epsfig, color}
\usepackage{amssymb}
\usepackage{amsmath}
\usepackage{hyperref}
\usepackage{bm}
\usepackage{cleveref}
\usepackage{braket}
\begin{document}
\title{Polaronic Response of a Supersonic Impurity Strongly Coupled to a Bose Condensate}
\author{Sooshin Kim}\thanks{These authors contributed equally to this work.}
\author{Yoonsoo Kim}\thanks{These authors contributed equally to this work.}
\author{Seokmin Jang}
\author{Jee Woo Park}\email{jeewoopark@postech.ac.kr}
\affiliation{Department of Physics, Pohang University of Science and Technology (POSTECH), Pohang 37673, Korea}

\begin{abstract}
How a mobile impurity exchanges momentum and energy with a many-body environment is a central question in nonequilibrium quantum physics. This exchange becomes particularly complex when the impurity moves rapidly and is strongly coupled to its bath. Here, we realize an interaction-tunable cold-atom collider with fermionic $^{40}$K impurities immersed in a $^{23}$Na Bose--Einstein condensate. A species-selective Raman pulse simultaneously launches the condensate and quenches the interspecies scattering length, producing an initial relative speed 36 times the condensate speed of sound. We track the ensuing relative motion as the interspecies interaction is tuned from weak coupling to resonance. At weak and intermediate coupling, the impurity dynamics are well described by a finite-energy two-body collision model. Near resonance, however, the early-time impurity acceleration is markedly suppressed relative to the two-body prediction, corresponding to a strong enhancement of the apparent dynamical inertia. The enhancement is observed near two distinct Feshbach resonances. These observations provide evidence for a polaronic response in the strongly coupled, supersonic regime of a degenerate Bose--Fermi mixture.
\end{abstract}

\maketitle

Scattering of a mobile particle from a quantum medium provides a fundamental probe of its excitations. In the weak-coupling limit, the probability for the projectile to transfer energy and momentum is governed by the dynamical correlation functions of the unperturbed medium, allowing the projectile to be treated as a passive probe within linear response~\cite{1954VanHove, 1978Squires_Book}. This separation ceases to be valid when the coupling becomes sufficiently strong that the projectile appreciably modifies the medium. Dressing by the medium's excitations then renormalizes the impurity's motion, turning the scattering process into the nonequilibrium dynamics of a coupled impurity--bath system~\cite{2025GrusdtDemlerArdila_ReviewBosePolaron}.

The polaron is the canonical example of such impurity dressing~\cite{1933Landau, 1946Pekar, 1954Frohlich, 1955Feynman}. In ultracold quantum gases, an impurity immersed in a Bose--Einstein condensate (BEC) becomes dressed by condensate excitations, forming a Bose polaron with an impurity--bath interaction that can be precisely tuned across a Feshbach resonance~\cite{2010ChinGrimmJulienne_Review}. Experiments have characterized Bose-polaron spectra and lifetimes~\cite{2016Jin_BosePolaron, 2016Arlt}, tracked their formation after an interaction quench~\cite{2021Skou_BosePolaronFormation}, and explored universal dynamics~\cite{2025Etrych_UniversalBosePolaronDynamics}, finite-temperature response~\cite{2020YanZwierlein_PolaronCriticality}, and low-energy collective motion~\cite{2024YanDemlerZwierlein_BreathingPolaron}. These studies have predominantly involved impurities prepared at rest or with little directed momentum relative to the bath.

Introducing a finite relative momentum between the impurity and the bath accesses a qualitatively different dynamical regime. At zero temperature, a dressed impurity can propagate without steady drag below the Landau critical velocity, whereas above it, the Cherenkov-like emission of Bogoliubov excitations transfers energy and momentum to the condensate~\cite{1998Cote, 2004Pitaevskii, 2006Carusotto}. At strong coupling, the impurity dispersion and the critical momentum for Cherenkov emission are themselves renormalized by dressing, and after a sudden interaction quench, the buildup of correlations occurs concurrently with momentum relaxation~\cite{2021Demler, 2024Demler}. High-velocity impurity scattering has been studied at weak coupling in three-dimensional BECs~\cite{2000Ketterle_Raman}, while interaction-tunable collisions have been investigated in thermal gases~\cite{2017Niels_AODCollider}, and finite-momentum relaxation at strong coupling has recently been explored in one dimension~\cite{2026Horvath_DissipationlessImpurity, 2025Petkovic_MobileImpurityin1DBoseGas}. However, the interaction-resolved dynamics of a highly supersonic impurity near resonance in a three-dimensional BEC remain largely unexplored.

In this Letter, we study the interaction-resolved scattering dynamics of fermionic impurities moving at highly supersonic velocities relative to a weakly interacting BEC. To the Bose--Fermi mixture, we apply a species-selective Raman pulse that simultaneously imparts a large two-photon recoil momentum to the condensate and transfers it into the hyperfine state whose interaction with the impurities is resonantly tunable. The pulse thereby quenches the interspecies scattering length and prepares an initial relative speed approximately 36 times the speed of sound of the condensate, driving the system deep into the dissipative Cherenkov regime. To probe the ensuing drag dynamics, we allow the two species to interact for a variable duration and repeat this measurement over a broad range of interspecies scattering lengths, resolving how the impurity momentum relaxation evolves from weak coupling to resonance. At weak and intermediate coupling, the measured dynamics are well described by a semiclassical two-body collision model, establishing a quantitative baseline for the experiment. Near resonance, analysis with the same framework reveals a marked increase in the apparent dynamical inertia of the impurity, suggesting that dynamical many-body dressing modifies the impurity response in the strongly coupled, supersonic regime.

\begin{figure}
	\centering
	\includegraphics{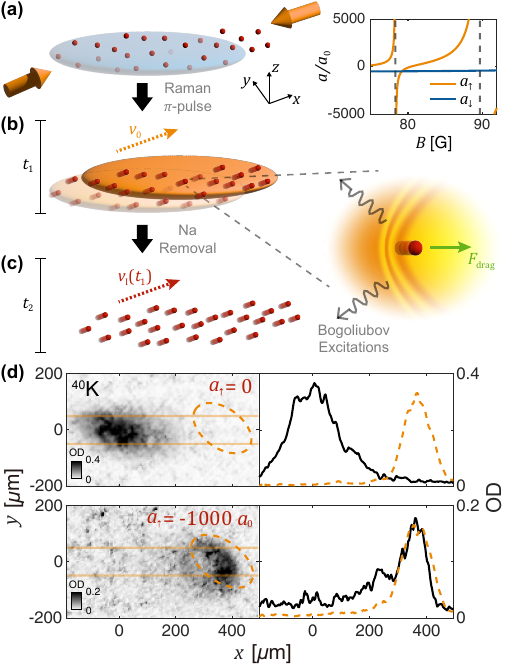}
	\caption{Interaction-tunable cold-atom collider. (a) Fermionic $^{40}$K impurities (red) immersed in a highly oblate $^{23}$Na BEC initially prepared in $\ket{\downarrow}$ (blue). A Raman $\pi$ pulse transfers the condensate to $\ket{\uparrow}$ (orange), imparting a lab-frame velocity $v_0 \hat{\mathbf{x} }$ by the two-photon recoil and quenching the Na--K scattering length from $a_{\downarrow}$ to $a_{\uparrow}$. The right inset shows the magnetic-field dependence of $a_{\downarrow}$ and $a_{\uparrow}$; gray dashed lines mark the positions of two neighboring interspecies Feshbach resonances. (b) During the interaction time $t_1$, the moving Na bath transfers momentum to the K impurities, which acquire the lab-frame velocity $v_\mathrm{I}(t_1)\hat{\mathbf{x} }$. The magnified schematic depicts the drag force $F_\mathrm{drag}$ on an impurity and the accompanying Cherenkov-like emission of Bogoliubov excitations. (c) After $t_1$, the Na bath is removed and the K cloud evolves ballistically for a duration $t_2$. (d) K-absorption images in the $xy$ plane and corresponding $y$-averaged $x$ profiles at $t_1 = 5\,\mathrm{ms}$ for $a_{\uparrow} = 0$ (top) and $- 1000a_0$ (bottom), each averaged over five realizations. For these measurements, the Na bath is retained and $t_2 = 0$. Horizontal yellow lines delimit the $y$-averaging window. Yellow dashed contours mark the BEC Thomas--Fermi boundaries at the same moment, while dashed curves show the corresponding condensate profiles rescaled vertically for clarity.}
	\label{Figure1}
\end{figure}

The experimental sequence is illustrated in Fig.~\ref{Figure1}(a)--(c). We begin with a quantum-degenerate mixture of $^{23}$Na bosons and $^{40}$K fermions confined in a highly oblate optical dipole trap (ODT)~\cite{2026Park_Machine, SM} [Fig.~\ref{Figure1}(a)]. A typical sample contains $1.7(1) \times 10^5$ Na atoms with a condensate fraction of 0.56(1), together with $2.3(2) \times 10^4$ K atoms at a temperature $T = 64(14)\,\mathrm{nK}$, corresponding to $T / T_\mathrm{F} = 0.5(1)$, where $T_\mathrm{F} = 123(3)\,\mathrm{nK}$ is the Fermi temperature of the trapped K gas. Their mean densities are $n_{\rm B} = 9.9(3) \times 10^{12}\,\text{cm}^{-3}$ for the Na condensate and $n_{\rm I} = 1.01(4) \times 10^{11}\,\text{cm}^{-3}$ for the K impurities, corresponding to a mean impurity concentration $n_{\rm I} / n_{\rm B} \simeq 1.02(5)\%$. The Na and K atoms are initially prepared in the hyperfine states $\ket{F, m_F} = \ket{1, 0} \equiv \ket{\downarrow}$ and $\ket{9 / 2, - 9 / 2}$, respectively. We then ramp the magnetic field into the region between two neighboring interspecies Feshbach resonances for collisions between Na $\ket{1, 1} \equiv \ket{\uparrow}$ and K $\ket{9 / 2, - 9 / 2}$~\cite{2012ParkWillZwierlein, 2016Simoni}. Over this field range, the scattering length $a_{\uparrow}$ between Na $\ket{\uparrow}$ and K is widely tunable, whereas the initial-state scattering length $a_{\downarrow}$ between Na $\ket{\downarrow}$ and K remains small and nearly field-independent~\cite{SM}. During the subsequent magnetic-field stabilization, we adiabatically turn on a K-selective optical guide propagating along the $x$-axis. The guide provides transverse confinement for the K atoms and allows us to adjust the relative position between the clouds of the two species to improve their spatial overlap~\cite{Guide_Beam}.

To launch the Na bath while simultaneously quenching its interaction with K impurities, we drive a transition from the Na $\ket{\downarrow}$ to the $\ket{\uparrow}$ state using a pair of counterpropagating Raman beams along the $x$-axis. The $\pi$ pulse has a duration of $4.68(2)\,\mu\mathrm{s}$, shorter than the mean interspecies collision time~\cite{SM}, and it simultaneously imparts a two-photon recoil momentum to the Na atoms [Fig.~\ref{Figure1}(b)]. The pulse thus effectively quenches the interspecies scattering length from $a_{\downarrow}$ to the target value of $a_{\uparrow}$, while launching the Na bath at the laboratory-frame velocity $v_0 = 2 \hbar k_\mathrm{R} / m_\mathrm{B} = 58.9\,\mathrm{mm/s}$ along $+ \hat{\mathbf{x} }$, where $\hbar = h / 2 \pi$ is the reduced Planck's constant, $k_\mathrm{R}$ is the wavenumber of the Raman beams, and $m_\mathrm{B}$ is the Na atomic mass. Note that the Raman beams address the Na bath alone without disturbing the K impurities, as their frequencies lie near the Na $D$ lines~\cite{SM}; $v_0 \hat{\mathbf{x} }$ therefore sets the initial relative velocity between the two species.

\begin{figure*}[!t]
	\centering
	\includegraphics{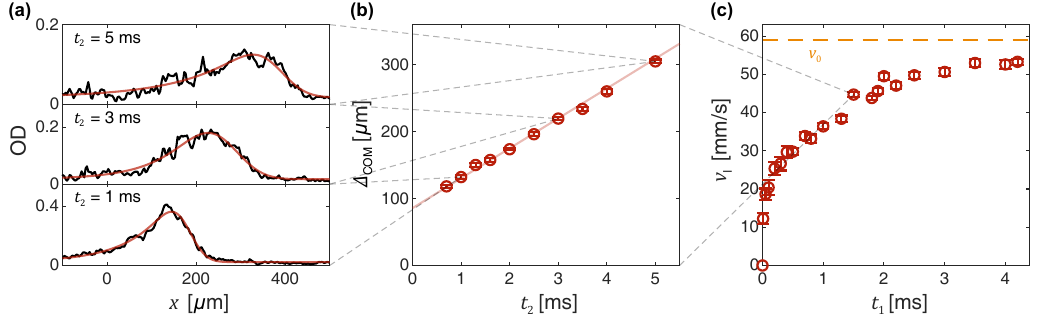}
	\caption{Reconstruction of the impurity velocity. (a) One-dimensional impurity-density profiles at $t_2 = 1, 3, 5\,\mathrm{ms}$ (bottom to top) for $t_1 = 1.5\,\mathrm{ms}$ and $a_{\uparrow} / a_0 = - 1000$. Each profile is obtained from a single experimental realization; solid red curves show the fits~\cite{SM}. (b) Impurity-COM displacement $\Delta_\mathrm{COM}$ as a function of $t_2$. Each point is the mean of five realizations, with error bars denoting the $1\sigma$ standard error of the mean. The solid line is a linear fit whose slope yields the laboratory-frame impurity speed $v_\mathrm{I}(t_1)$. Note that the nonzero $y$-intercept reflects the impurity position at the end of $t_1$. (c) Reconstructed $v_\mathrm{I}(t_1)$ for different interaction times $t_1$. The yellow dashed line marks the speed of the bath $v_0 = 58.9\,\mathrm{mm/s}$. Error bars denote the $1\sigma$ uncertainties of the corresponding fitted slopes.}
	\label{Figure2}
\end{figure*}

The ensuing scattering dynamics can be described in the frame comoving with the Na condensate, where the K impurities are initialized with average velocity $- v_0 \hat{\mathbf{x} }$. At the mean condensate density $n_\mathrm{B}$, the Bogoliubov speed of sound is $c = (4 \pi \hbar^2 a_\mathrm{BB} n_\mathrm{B} / m_\mathrm{B}^2)^{1 / 2} = 1.62(2)\,\mathrm{mm / s}$, where $a_\mathrm{BB}$ is the Na--Na scattering length. The ratio $v_0 / c = 36.3(5)$ places the impurities deep in the supersonic regime. For a nonzero $a_{\uparrow}$, Cherenkov-like emission of Bogoliubov excitations produces a drag force opposite to the impurity motion, thereby reducing the relative speed~\cite{1999Ketterle, 2001Maksimov, 2004Pitaevskii, 2006Carusotto, 2021Demler, SM}. In the laboratory frame, the same drag force accelerates the K impurities in the direction of the moving Na bath. We allow the two species to interact for a variable duration $t_1$~\cite{Escape_ODT}, after which the K cloud has acquired the laboratory-frame center-of-mass (COM) velocity $v_\mathrm{I}(t_1) \hat{\mathbf{x} }$.

At the end of $t_1$, we remove the Na bath using a $100\,\mu\mathrm{s}$ resonant light pulse and then switch off the ODT, allowing the K cloud to ballistically evolve along $\hat{\mathbf{x} }$ for a variable duration $t_2$ before absorption imaging [Fig.~\ref{Figure1}(c)]. We determine $v_\mathrm{I}(t_1)$ from the linear dependence of the K-COM position on $t_2$. To visualize the interaction-induced motion directly, representative K absorption images acquired with the Na bath retained and $t_2 = 0$ are shown in Fig.~\ref{Figure1}(d). At $t_1 = 5\,\mathrm{ms}$, the K cloud remains near its initial position for $a_{\uparrow} = 0$, whereas for $a_{\uparrow} = - 1000 a_0$, it is displaced in the direction of the moving bath, with $a_0$ being the Bohr radius. This contrast provides a direct qualitative signature of momentum transfer from the bath to the impurities.

Figure~\ref{Figure2} shows how the impurity velocity is reconstructed quantitatively. For each fixed interaction strength $a_{\uparrow}$ and interaction time $t_1$, we record one-dimensional impurity profiles over a range of $t_2$ [Fig.~\ref{Figure2}(a)] and extract their COM positions~\cite{SM}. The corresponding displacement $\Delta_\mathrm{COM}$ grows linearly with $t_2$, as displayed in Fig.~\ref{Figure2}(b). Here, the fitted slope yields the laboratory-frame impurity speed $v_\mathrm{I}(t_1)$. Repeating this procedure over a range of $t_1$ reconstructs the kinetic evolution shown in Fig.~\ref{Figure2}(c), with each point obtained from an independent scan over $t_2$. For the representative interaction shown, $v_\mathrm{I}(t_1)$ initially rises rapidly and then approaches a plateau near the speed of the bath $v_0$. As $d v_\mathrm{I} / d t_1$ gives the impurity-COM acceleration, the resulting velocity traces provide direct experimental access to the interaction-dependent drag response.

To analyze the drag dynamics quantitatively, we start from Fermi's golden rule at weak impurity--bath coupling, which expresses the momentum-transfer rate in terms of the dynamic structure factor of the condensate~\cite{1998Cote}. In the highly supersonic limit, the Bogoliubov suppression of scattering becomes negligible (here $\frac{\mu v_0}{\hbar} \xi \approx 16 \gg 1$, where $\mu = m_\mathrm{I} m_\mathrm{B} / (m_\mathrm{I} + m_\mathrm{B} )$ is the impurity--bath reduced mass with $m_\mathrm{I}$ being the K atomic mass and $\xi \approx 1.2\,\mu\mathrm{m}$ the condensate healing length), and the golden-rule result reduces to a simpler kinetic picture of independent binary collisions between the impurity and condensate atoms~\cite{SM}. In the frame of the bath, an impurity with cross section $\sigma$ moving at relative speed $u$ through a condensate of density $n_\mathrm{B}$ sweeps out a collision volume $\sigma u$ per unit time, providing the collision rate $\Gamma = n_\mathrm{B} \sigma u$. Within this collision picture, $\sigma = 4 \pi a_{\uparrow}^2/(1 + k_\mathrm{col}^2 a_{\uparrow}^2)$ is the finite-energy elastic $s$-wave scattering cross section, where $k_\mathrm{col} = \mu u / \hbar$ is the relative collision wavenumber~\cite{2011Zwierlein, SM}. As the mean longitudinal momentum transfer is given to be $\mu u$ per collision, the drag force is then
\begin{equation}
	F_\mathrm{drag} = \mu u \Gamma = \frac{4 \pi \hbar^2 n_\mathrm{B} }{\mu} \frac{\eta^2}{1 + \eta^2},
	\label{eq:Drag_Force}
\end{equation}
where $\eta = k_\mathrm{col} a_{\uparrow}$ is the dimensionless collision parameter.

We obtain the predicted velocity evolution by integrating the impurity equation of motion, assuming a homogeneous bath of constant density $n_\mathrm{B}$ and neglecting bath recoil. With $u(t_1) = v_0 - v_\mathrm{I}(t_1)$, the equation of motion reads $F_\mathrm{drag} = - m_\mathrm{I} \frac{du}{dt_1} = - \frac{\hbar m_\mathrm{I} }{\mu a_{\uparrow} } \frac{d \eta}{d t_1}$. Substituting Eq.~\eqref{eq:Drag_Force} and integrating from the initial condition $\eta_0 = \mu v_0 {a_{\uparrow} } / \hbar$ results in
\begin{equation}
	Y(t_1) \equiv \frac{m_\mathrm{I} }{4 \pi \hbar n_\mathrm{B} a_{\uparrow} } \left[\eta_0 - \eta(t_1) + \frac{1}{\eta(t_1) } - \frac{1}{\eta_0} \right] = t_1.
	\label{eq:Linearized_Response}
\end{equation}
Thus, within this collision model, the transformed response $Y(t_1)$ is linear in $t_1$ with unit slope.

\begin{figure}
	\centering
	\includegraphics{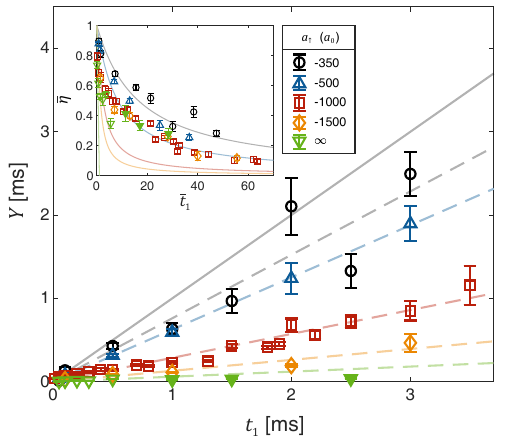}
	\caption{Response of impurities to the drag force for various interaction strengths. In the inset, we plot the normalized relative speed $\bar{\eta}(\bar{t}_1)$ as a function of the dimensionless time $\bar{t}_1 \equiv \frac{4 \pi \hbar^2 n_\mathrm{B} }{m_\mathrm{I} \mu v_0} t_1$. Solid lines are expectation from solving equation of motion with the drag force Eq.~\eqref{eq:Drag_Force}. The main panel presents the same data transformed to a linearized response $Y(t_1)$. For any scattering length $a_{\uparrow}$, the slope of this plot is anticipated to converge to unity (gray solid line) in the two-body collision model discussed. Dashed lines are linear fit to the data~\cite{Fit_Range}. Error bars denote $1\sigma$ statistical uncertainties.}
	\label{Figure3}
\end{figure}

Figure~\ref{Figure3} shows how the impurity dynamics evolve as the interaction is increased toward unitarity. The inset shows the evolution of the normalized relative speed $\bar{\eta} \equiv \eta / \eta_0$ for representative interaction strengths. As expected, the relative speed decreases more rapidly at stronger interactions. This overall trend, however, masks a substantial deviation from the two-body prediction near resonance. To quantify this deviation, we transform the same data according to Eq.~\eqref{eq:Linearized_Response}. The resulting $Y(t_1)$ traces in the main panel remain approximately linear over the measured time range for all displayed interactions, even in the strongly interacting regime, allowing each trace to be characterized by a fitted slope $s(a_{\uparrow} ) = dY / dt_1$. Note that the prediction from the collision model is $s = 1$. However, in our experiment, $s(a_{\uparrow} )$ decreases markedly as $|a_{\uparrow}|$ increases, indicating that the measured acceleration of the impurity is increasingly suppressed relative to what the two-body model predicts.

\begin{figure}
	\centering
	\includegraphics{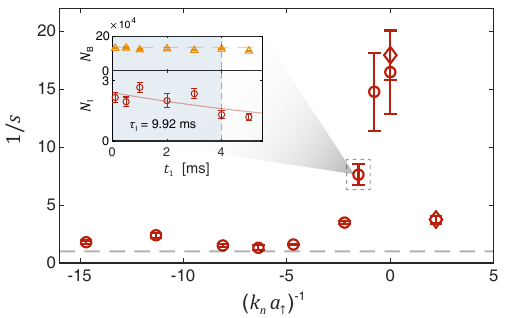}
	\caption{Interaction dependence of the dynamical inertia. The inverse slope $\frac{1}{s}$ extracted from the linearized response is plotted as a function of $(k_n a_{\uparrow})^{-1}$. Circles and diamonds denote measurements performed near the lower- and higher-field Feshbach resonances, respectively~\cite{SM}. The gray dashed line marks the two-body baseline $s = 1$. The inset shows the atom-number loss of the bath ($N_\mathrm{B}$, yellow) and impurity ($N_\mathrm{I}$, red) at $a_{\uparrow} = - 1500 a_0$ as functions of $t_1$. The yellow dashed line is the mean initial bath-atom number of this data point. The red solid line is a simple exponential fit to the impurity-atom number, which yields the $1 / e$-lifetime of the impurities $\tau_\mathrm{I} = 9.92\,\mathrm{ms}$. For this data point, we used 4\,ms as a maximum $t_1$ (indicated as a blue-shaded area) to obtain the fitted slope $s$. Error bars denote the $1 \sigma$ statistical uncertainties.}
	\label{Figure4}
\end{figure}

Figure~\ref{Figure4} reveals the full interaction dependence of the fitted slopes, including data for scattering lengths not shown in Fig.~\ref{Figure3}.
The plotted inverse slopes measure the suppression of the impurity acceleration relative to the force scale predicted by the collision model. According to Eq.~\eqref{eq:Linearized_Response}, these values can be interpreted as an apparent dynamical inertia $m_\mathrm{app} / m_\mathrm{I} \equiv \frac{1}{s(a_{\uparrow} ) }$, as long as the effective condensate density sampled by the impurity and the underlying drag-force law remain unchanged.
We plot this as a function of $(k_n a_{\uparrow} )^{-1}$, where $k_n = (6 \pi^2 n_\mathrm{B} )^{1 / 3}$ is the condensate-density wavenumber. The inferred inertia remains nearly constant and close to the bare-mass expectation over the weak- and intermediate-coupling regimes, but rises sharply near unitarity, reaching $m_\mathrm{app} / m_\mathrm{I} \simeq 18$. This near-unitarity enhancement is reproduced around both neighboring resonances, including on the repulsive side, indicating that this physics is not restricted to a specific Feshbach resonance or to attractive interactions.

Because the collision-model force is proportional to $n_\mathrm{B}$, a reduction of the condensate density would also increase $\frac{1}{s}$. However, the interaction-dependent losses measured with the bath retained and $t_2 = 0$ are far too small to account for the observed increase. For example, in the inset of Fig.~\ref{Figure4}, the atom-number loss at $a_{\uparrow} = - 1500 a_0$, where $\frac{1}{s}$ already deviates significantly from the baseline, is shown as a function of $t_1$. Since $n_\mathrm{B} \gg n_\mathrm{I}$, the enhanced loss at large $|a_{\uparrow}|$ is registered predominantly in the impurity number rather than in that of the bath, and the corresponding reduction of $n_\mathrm{B}$ is small. The Na cloud widths also remain nearly constant over the relevant times~\cite{SM}. Moreover, we extract the fitted slopes from the early-time dynamics to limit the influence of accumulated atom loss~\cite{SM}. The measured atom loss is therefore insufficient to account for the sharp increase near unitarity through bulk condensate depletion.

A polaronic enhancement of the dynamical impurity inertia provides a natural interpretation of this increase. At strong coupling, the impurity atom becomes correlated with a cloud of condensate excitations, so that changing its velocity involves the response of a composite impurity--bath object rather than that of the bare atom alone. An approximately time-independent enhancement of this dynamical inertia would preserve the near-linear $Y(t_1)$ traces while reducing their slopes, as observed. The same many-body dressing may also produce a transient redistribution of the condensate density in the vicinity of the moving impurity, further modifying the local drag response~\cite{2019MistakidisGrusdtSchmelcher}. Note that since the experiment probes a strongly dissipative nonequilibrium state~\cite{2019Pohl}, $m_\mathrm{app}$ need not coincide quantitatively with the equilibrium effective mass of a Bose polaron~\cite{2012Catani, 2013Schmidt, 2014DasSarma} defined in a stationary system. Nevertheless, the sharp, resonance-localized increase provides evidence for the onset of a many-body enhanced dynamical inertia of the supersonic impurity.

In summary, we have realized an interaction-tunable cold-atom collider in which a species-selective Raman pulse simultaneously launches a $^{23}$Na condensate and quenches its interaction with $^{40}$K impurities. At weak and intermediate coupling, the measured velocity evolution is quantitatively captured by a finite-energy binary-collision model. Near resonance, however, the impurity acceleration is strongly suppressed relative to this prediction, which can be interpreted as a pronounced increase of the apparent dynamical inertia. The enhancement is reproduced near both Feshbach resonances and cannot be accounted for by the condensate density reduction inferred from the measured atom loss. These observations have implications that many-body dressing modifies the response of a deeply supersonic impurity and support its interpretation as a polaronic enhancement of the impurity's dynamical inertia in a regime where dressing and Cherenkov dissipation develop concurrently. 

Looking forward, the independent control of interaction strength and injected momentum opens a direct route to mapping the crossover between subsonic polaron motion and Cherenkov dissipation~\cite{2021Demler, 2024Demler}. Specifically, reducing the initial speed to near $c$ should allow us to microscopically explore the interplay between polaron formation and dissipative relaxation~\cite{2023Fleischhauer}. Such measurements would also provide a stringent benchmark for beyond-mean-field treatments of finite-momentum polaron dynamics at arbitrary impurity--boson couplings~\cite{2025GrusdtDemlerArdila_ReviewBosePolaron}. Extending the collider to a near-homogeneous confinement and combining it with spectroscopic and spatially resolved probes could connect the enhanced dynamical inertia to the equilibrium effective mass of the Bose polaron~\cite{2015Bruun_MobileImpurity, 2025Moritz} and to the finite-momentum quasiparticle spectrum, recently mapped out for the Fermi polaron~\cite{2026GrimmBruun_MovingFermiPolaron}.

\begin{acknowledgments}
We acknowledge helpful discussions with Y.-i.~Shin and M.~Fleischhauer. We thank D. Kwon for experimental assistance. This work was supported by the National Research Foundation of Korea (NRF) (Grant Nos. RS-2025-00559423, RS-2025-02317602, RS-2025-25456929, RS-2023-NR119928), the Samsung Science and Technology Foundation (No. SSTFBA2001-06), and the Ministry of Science and ICT (MSIT), Korea, under the Information Technology Research Center (ITRC) support program (No. RS-2022-00164799) supervised by the Institute of Information \& Communications Technology Planning \& Evaluation (IITP). S.~Kim acknowledges support from the NRF grant funded by the MSIT (No. RS-2024-00394359) and the Basic Science Research Institute Fund (No. 2021R1A6A1A10042944). Y.~Kim and S.~Jang acknowledge support from the Basic Science Research Program through the NRF funded by the Ministry of Education (Nos. RS-2024-00410492 and RS-2025-25421720, respectively).
\end{acknowledgments}

\bibliography{References}

\appendix

\clearpage

\section{SUPPLEMENTAL MATERIAL}

\subsection{Calibration of parameters}

To calibrate the Na trapping frequencies of the 1064\,nm optical dipole trap (ODT), we excite a center-of-mass (COM) oscillation of a Na Bose--Einstein condensate (BEC) and record the subsequent motion by absorption imaging after time-of-flight (TOF) expansion. The measured eigenfrequencies are $2 \pi \times (10.30(1), 5.58(3), 560(4))\,\mathrm{Hz}$. The strongest confinement is along $\hat{\mathbf{z}}$, while the weakest is along the ODT beam direction, $- \frac{\sqrt{3}}{2} \hat{\mathbf{x} } - \frac{1}{2} \hat{\mathbf{y} }$, so that the two in-plane eigenaxes are rotated with respect to $\hat{\mathbf{x} }$ and $\hat{\mathbf{y} }$. For $^{40}$K, we obtain the eigenfrequencies from parametric heating, which yields $2 \pi \times (20.3(2), 8.7(1), 653(2))\,\mathrm{Hz}$ for the combined trap of the ODT and the K-selective optical guide. Applying the same method to Na in the ODT alone reproduces the frequencies obtained from the COM oscillation.

We calibrate the magnetic field, which sets the interspecies scattering length $a_{\uparrow}$, using two-photon Raman spectroscopy between the $|1, 0 \rangle$ and $|1, 1 \rangle$ hyperfine states of Na. For each target field, we set the frequency difference between the counterpropagating Raman beams to the transition frequency calculated from the Breit--Rabi formula, reduced by the two-photon recoil frequency of $100\,\mathrm{kHz}$. We then scan the Helmholtz-coil current and locate the transfer resonance, obtaining a magnetic-field uncertainty of $\sim 5\,\mathrm{mG}$, which is negligible on the scale of the horizontal axis of Fig.~\ref{Figure4} of the main text.

\subsection{Experimental sequence}

In our experiment, we first prepare a quantum-degenerate Bose--Fermi mixture, with $^{23}$Na condensed in the $|1, 1 \rangle$ hyperfine state and $^{40}$K in $|9 / 2, 9 / 2 \rangle$~\cite{2026Park_Machine}, trapped in a highly oblate ODT formed by a focused 1064\,nm beam~\cite{2024Park}. Over the full range of interaction strengths explored, the Na and K atom numbers at the onset of the dynamics show no systematic dependence on $a_{\uparrow}$. For Na, the condensate density $n_\mathrm{B}$ remains approximately constant, with a maximum deviation below 11\%.

Subsequently, we transfer Na to $|1, 0\rangle \equiv |{\downarrow}\rangle$ and K to $|9 / 2, - 9 / 2\rangle$ using Landau--Zener sweeps at a bias field of $2.35\,\mathrm{G}$. The field is then ramped along $\hat{\mathbf{z} }$ over 25\,ms to a target value $B_z$ in the range 78.3--89.7\,G, bounded by two neighboring Feshbach resonances for collisions between Na $|1, 1\rangle \equiv |{\uparrow}\rangle$ and K $|9 / 2, - 9 / 2\rangle$~\cite{2012ParkWillZwierlein, 2016Simoni}. Within this range, the initial-state scattering length $a_{\downarrow}$ is nearly field-independent, as the Na $|{\downarrow}\rangle$--K mixture shows no measurable atom loss over 2\,s, far from any Feshbach resonance (see the following section).

Over the next 1\,s, we stabilize the magnetic field and adiabatically turn on the K-selective optical guide. The guide is formed by a weakly focused 775\,nm beam, close to the K $D$ lines, propagating along $+ \hat{\mathbf{x} }$ with a waist of $\sim 165\,\mu\text{m}$ and a power of 48.6\,mW. Balancing this potential against a residual magnetic-field gradient, we place most of the K atoms slightly ahead of the Na cloud along $+ \hat{\mathbf{x} }$, so that the condensate sweeps through the impurities once it is launched in that direction. The guide also narrows the K cloud in the transverse directions to widths comparable to the corresponding Thomas--Fermi radii of the condensate, improving their transverse overlap, as illustrated in Fig.~\ref{Figure1}(d) of the main text. Its longitudinal trapping frequency is about $2 \pi \times 0.06\,\mathrm{Hz}$, with an estimated fractional correction on the order of $10^{-6}$ to the ballistic K-COM displacement over the longest $t_2$.

\begin{figure}
	\centering
	\includegraphics{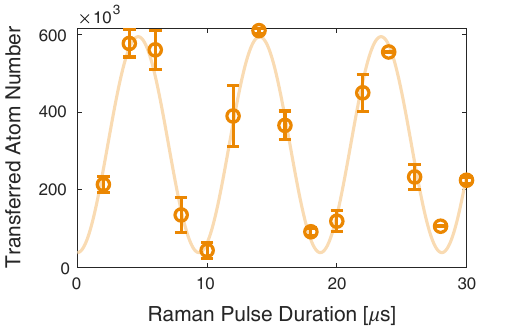}
	\caption{Rabi oscillation between the $| {\downarrow} \rangle | \mathbf{p} = 0 \rangle$ and $| {\uparrow} \rangle | \mathbf{p} = 2 \hbar k_\mathrm{R} \hat{\mathbf{x} } \rangle$ states of Na, driven by a pair of counterpropagating Raman beams, as a function of pulse duration. Here, $\mathbf{p}$ is the momentum of a Na atom. The solid line is a sinusoidal fit to the measured number of transferred atoms, from which we obtain the $\pi$-pulse duration quoted in the text. The measurement is performed on a Na-only sample, with no K atoms loaded. Error bars denote $1\sigma$ statistical uncertainties.}
	\label{FigureS_Raman_Rabi}
\end{figure}

We then transfer the Na atoms from $|{\downarrow}\rangle$ to $|{\uparrow}\rangle$ using a short Raman $\pi$ pulse. Raman beams A and B, each of $\sim 176\,\text{mW}$, propagate along the $+ \hat{\mathbf{x} }$ and $- \hat{\mathbf{x} }$ directions and are linearly polarized along the $y$ and $z$ axes, respectively. With the quantization axis along $\hat{\mathbf{z} }$, the transition is driven by the $\sigma^{+}$ component of beam A and the $\pi$ polarization of beam B. Both beams are collimated to a $1 / e^2$ diameter of $\sim 1.4\,\text{mm}$. The frequencies of the Raman beams lie between the Na $D_1$ and $D_2$ lines near a tune-out condition, approximately $172\,\text{GHz}$ above the $D_1$ resonance, suppressing the optical dipole force on Na while retaining strong two-photon coupling. The beams are far detuned from the K transitions and leave the impurities unaffected. Here, the Raman B beam has a higher frequency than the Raman A beam, so the Na atoms absorb photons from the Raman A beam and undergo stimulated emission into the Raman B beam, thereby gaining momentum toward $+ \hat{\mathbf{x} }$. As shown in Fig.~\ref{FigureS_Raman_Rabi}, we observe coherent Rabi oscillations over multiple cycles and obtain a $\pi$-pulse duration of $4.68(2)\,\mu\text{s}$. This is shorter than the mean interspecies collision time at unitarity, $\tau_\mathrm{col} = (n_\mathrm{B} \sigma_\mathrm{u} v_0)^{-1} \approx 25\,\mu\mathrm{s}$, where $\sigma_\mathrm{u} = 4 \pi / k_\mathrm{col}^2$ is the unitarity-limited cross section, and more than two orders of magnitude shorter at the weakest coupling explored. The pulse therefore acts as a sudden quench of both the interaction and the relative velocity.

The mixture is then allowed to interact for a variable duration $t_1$, up to 6\,ms. The usable range of $t_1$ depends on $a_{\uparrow}$. An excessively long $t_1$ becomes uninformative once the relative displacement of the two clouds exceeds the size of the condensate. An excessively short $t_1$ does not impart a momentum change large enough for the drag contribution to dominate over other effects, such as residual magnetic-field gradients and the weak longitudinal force from the guide as gravity displaces the K atoms from its axis during $t_2$. For the values of $t_1$ retained in the analysis, $\Delta_\mathrm{COM}$ grows linearly with $t_2$ [Fig.~\ref{Figure2}(b)], confirming that these residual effects are negligible compared with the drag-induced motion. We also use only interaction times shorter than the measured lifetime at each coupling to fit $s(a_{\uparrow} )$. We further confirm experimentally that the atom numbers of a mixture with a finite relative velocity decay negligibly, except at the resonances, where the moving and stationary mixtures show similar lifetimes. We also verify that the relative widths of the Na cloud remain approximately constant for $t_1 < 10\,\mathrm{ms}$ by imaging in both the $xy$ and $zx$ planes. Taken together, these checks indicate that $n_\mathrm{B}$ remains nearly constant across all the experimental realizations in this work.

At the end of $t_1$, we remove all Na atoms using a $100\,\mu\mathrm{s}$ pulse along $+ \hat{\mathbf{z} }$ resonant with the Na $D_2$ transition, terminating the interspecies interaction. For efficient removal of Na, we find the resonant frequency for the Na $|{\uparrow}\rangle$ state at every value of $B_z$ used in the experiment. When characterizing the sample, as in Fig.~\ref{Figure1}(d) of the main text, we instead omit the pulse and retain the Na bath.

\begin{figure*}
	\centering
	\includegraphics{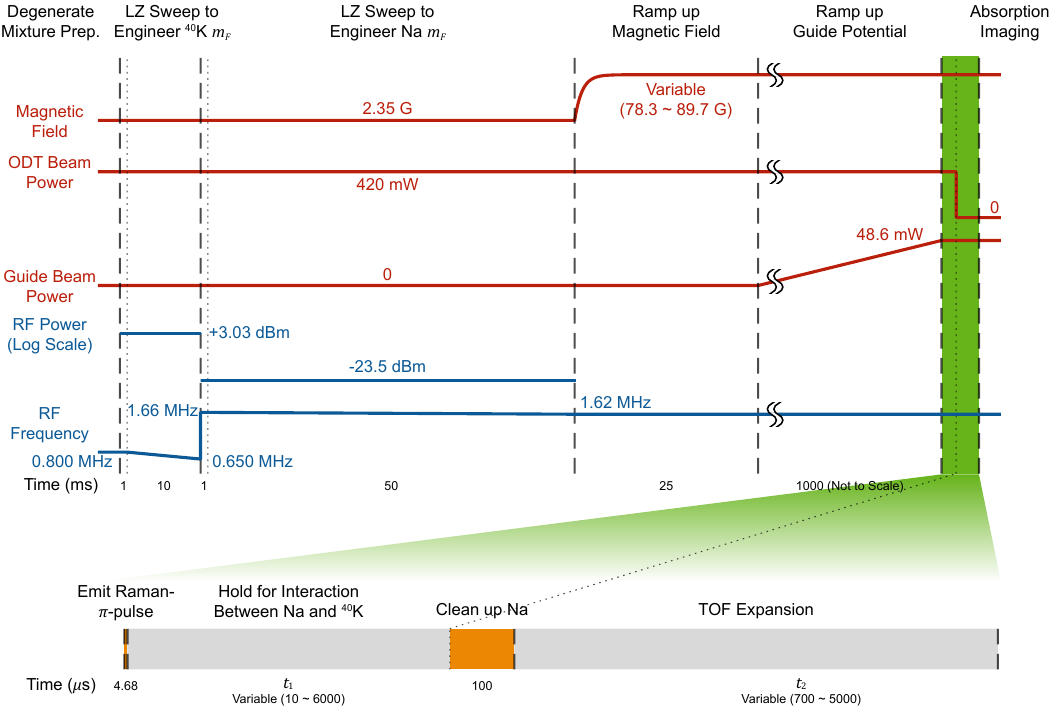}
	\caption{Experimental sequence. Major experimental parameters are shown as functions of time. Two frequency-swept RF pulses (Landau--Zener sweeps) are applied to prepare the desired $m_F$ states of $^{40}$K and Na at a low $B$ field. The system is then brought to a target magnetic field, and a guide potential acting predominantly on $^{40}$K is adiabatically switched on. A short Raman $\pi$ pulse is subsequently applied to initiate the dynamics, followed by a hold time $t_1$, removal of the Na atoms, and a TOF expansion for $t_2$. Finally, the $^{40}$K atoms are imaged using resonant absorption imaging.}
	\label{FigureS_Sequence}
\end{figure*}

The 1064\,nm ODT is then switched off, and the K cloud moves ballistically for a variable time $t_2$ at the velocity $v_\mathrm{I}(t_1)\hat{\mathbf{x} }$ acquired by the end of $t_1$. The K-selective optical guide remains on, but its longitudinal confinement is negligible, and the K-COM motion along $x$ is unaffected. We finally perform absorption imaging along $\hat{\mathbf{z} }$. The sequence is repeated over $t_2$ to obtain $v_\mathrm{I}(t_1)$, over $t_1$ to characterize the drag-induced acceleration, and over $B_z$ to map the interaction dependence of the drag response. Each combination of parameters is measured at least five times, for a total of more than 4000 shots. The detailed parameter ramps are shown in Fig.~\ref{FigureS_Sequence}.

To extract the COM coordinates, we first fit the one-dimensional profile along $x$ with an exponentially modified Gaussian to estimate the peak position of the skewed distribution. Using this peak position as an initial estimate, we perform a two-dimensional (2D) Gaussian fit to the original image. We then calculate the COM by taking a weighted sum over the pixels within the $2\sigma$ region defined by the Gaussian fit, reducing the contribution of noise outside the region of interest.

\subsection{Feshbach resonances between N\lowercase{a} $|F = 1, m_F = 0\rangle$ and $^{40}$K $|9 / 2, - 9 / 2\rangle$}

\begin{figure}
	\centering
	\includegraphics{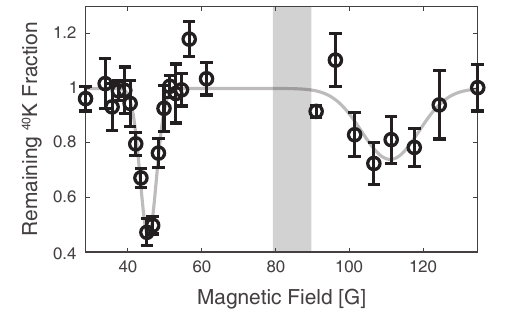}
	\caption{Experimental loss spectra of $^{40}$K $|9 / 2, - 9 / 2\rangle$ in the presence of $^{23}$Na $|1, 0\rangle$. The solid line shows Gaussian fits to the data. A $p$-wave resonance at 45.6(2)\,G with a width of 2.2(2)\,G and an $s$-wave resonance at 111(3)\,G with a width of 7(4)\,G are observed. The measurements are performed with a degenerate mixture of $\sim 1.2 \times 10^5$ atoms per each species. The shaded area represents the magnetic-field range used in the main experiment.}
	\label{FigureS_Na10_Resonances}
\end{figure}

While Feshbach resonances between Na $|{\uparrow}\rangle$ and $^{40}$K $|9 / 2, m_F\rangle$ are well explored both theoretically~\cite{2016Simoni} and experimentally~\cite{2012ParkWillZwierlein}, those between Na $|{\downarrow}\rangle$ and $^{40}$K $|9 / 2, m_F\rangle$ have not been reported. To characterize the initial-state channel used in our experiment, we perform Feshbach loss spectroscopy of Na $|{\downarrow}\rangle$ and $^{40}$K $|9/2,-9/2\rangle$, measuring the atom losses as a function of magnetic field. We focus on resonances closest to the magnetic-field range used in our experiment. The measured spectra are exhibited in Fig.~\ref{FigureS_Na10_Resonances}. We summarize their positions and widths in Table~\ref{TableS_Na10_Resonances}.

\begin{table}
	\centering
	\caption{Data summary on the Feshbach resonances between $^{23}$Na in $|1, 0\rangle$ and $^{40}$K in $|9 / 2, - 9 / 2\rangle$. We report here only the two resonances nearest to the range of the magnetic field employed in our experiment. The positions and widths of the resonances, $B_0^{\rm exp}$ and $\Delta B^{\rm exp}$, are determined by phenomenological Gaussian fits ($\propto e^{-(B - B_0^{\rm exp})^2 / (\Delta B^{\rm exp} )^2}$) to the observed loss features. $B_0^{\rm th}$ provides approximate theoretical positions from a simple asymptotic bound-state model.}
	\begin{ruledtabular}
		\begin{tabular}{cccc}
			$B_0^{\rm exp}$ [G] &
			$\Delta B^{\rm exp}$ [G] &
			$B_0^{\rm th}$ [G] &
			Resonance type \\
			\colrule
			45.6(2) & 2.2(2) & 31 & $p$ \\
			111(3) & 7(4) & 100 & $s$ \\
		\end{tabular}
	\end{ruledtabular}
	\label{TableS_Na10_Resonances}
\end{table}

\subsection{High-field imaging}

We image Na and K atoms directly at the magnetic field used to tune $a_{\uparrow}$. For Na $|1, 1\rangle$, we use a $\sigma^-$-polarized probe on the $D_2$ transition to the excited state adiabatically connected to $|F', m_F'\rangle = |0, 0\rangle$ in the weak-field limit. At our experimental fields, this transition is not fully closed, and optical pumping into off-resonant states makes the imaging efficiency dependent on the probe intensity and exposure time~\cite{2016Shin}. We use a resonant Gaussian probe beam with a peak intensity of $I_0 = 0.179\,\mathrm{mW}/\mathrm{cm}^2$ and an exposure time of $120\,\mu\mathrm{s}$. The measured imaging efficiency under these conditions is 0.49(2), which we use to calibrate the Na atom numbers reported in the main text. The same beam is also used for the Na-removal pulse. In contrast, K atoms in the stretched $|9/2,-9/2\rangle$ state support a closed $\sigma^-$ cycling transition to $|F',m_F'\rangle = |11/2,-11/2\rangle$ throughout the experimental field range.

\subsection{Derivation of drag force}

In the main text, we use the semiclassical binary-collision model of Eq.~\eqref{eq:Drag_Force} in the highly supersonic regime $u \gg c$, where Bogoliubov suppression of scattering is negligible. As $u$ approaches $c$, this suppression becomes important. Here, we derive a more general expression for the velocity-dependent drag force $F_\mathrm{drag}(u)$ using Fermi's golden rule and the Bogoliubov dynamic structure factor of a homogeneous, zero-temperature BEC.

For an impurity with wavevector $\mathbf{k} = m_\mathrm{I} \mathbf{u} / \hbar$ in the condensate rest frame, we denote the scattering rate from $\hbar \mathbf{k}$ to $\hbar(\mathbf{k} - \mathbf{q})$ by $\Gamma(\mathbf{k}; \mathbf{k} - \mathbf{q})$. Each scattering event transfers momentum $\hbar \mathbf{q}$ from the impurity to a Bogoliubov excitation. By axial symmetry, the transverse contributions cancel in the sum, and the drag force can be written as
\begin{equation}
	F_\mathrm{drag}(u) = \sum_{\mathbf{q}} \Gamma(\mathbf{k}; \mathbf{k} - \mathbf{q})\, \hbar q \cos\theta,
	\label{eq:General_Drag_Force}
\end{equation}
where $\theta$ is the angle between $\mathbf{q}$ and $\mathbf{k}$. 

In the weak-coupling limit, Fermi's golden rule relates the scattering rate to the dynamic structure factor of the BEC,
\[
    \Gamma(\mathbf{k};\mathbf{k}-\mathbf{q})
    = \frac{2\pi g_\mathrm{IB}^2}{\hbar V}
      s(\mathbf{q},\omega),
\]
where $g_\mathrm{IB}=2\pi\hbar^2a_{\uparrow}/\mu$ is the impurity--bath coupling constant, $\hbar\omega = \epsilon(\mathbf{k}) - \epsilon(\mathbf{k} - \mathbf{q})$ is the energy transferred to the condensate, with $\epsilon(\mathbf{k})=\hbar^2k^2/(2m_\mathrm{I})$, and $V$ is the normalization volume. For a dilute condensate at $T=0$, the Bogoliubov approximation gives~\cite{1998Cote}
\begin{equation}
	s(\mathbf{q}, \omega) \simeq \frac{n_\mathrm{B} q^2}{2 m_\mathrm{B} \omega_q} \delta(\omega - \omega_q),
	\label{eq:Structure_Factor}
\end{equation}
where $\omega_q = cq\sqrt{1 + q^2 / 16 \pi n_\mathrm{B} a_\mathrm{BB}}$ is the Bogoliubov dispersion of the BEC. Here, the delta function enforces energy conservation for the emission of a single Bogoliubov excitation. The drag therefore vanishes for $u \leq c$, consistent with the Landau criterion. For $u > c$, energy conservation sets a maximum excitation wavenumber $q_\mathrm{max}$. In the thermodynamic limit, we evaluate Eq.~\eqref{eq:General_Drag_Force} by substituting $\sum_{\mathbf{q}}\to V/(2\pi)^3\int d^3q$, with the radial integration restricted to $0\leq q\leq q_\mathrm{max}$.

The calculation above uses the zero-energy scattering cross section $\sigma_0 = 4 \pi a_{\uparrow}^2$. In our experiment, however, $|k_\mathrm{col}a_{\uparrow}|$ is not generally small, so the finite-energy dependence of the cross section must be retained. We therefore replace $\sigma_0$ with the elastic $s$-wave cross section obtained from the finite-energy scattering amplitude $1/f(k_\mathrm{col}) \approx -1/a_{\uparrow} - i k_\mathrm{col}$,
\begin{equation}
	\sigma(k_\mathrm{col}) = 4\pi |f(k_\mathrm{col})|^2 = \frac{4 \pi a_{\uparrow}^2}{1 + k_\mathrm{col}^2 a_{\uparrow}^2},
	\label{eq:Finite_Energy_Cross_Section}
\end{equation}
which reduces to $\sigma_0$ for $|k_\mathrm{col}a_{\uparrow}|\ll1$ and approaches the unitarity limit $4\pi/k_\mathrm{col}^2$ for $|k_\mathrm{col}a_{\uparrow}|\gg1$. For example, the initial relative collision wavenumber is $k_{\mathrm{col},0}=\mu v_0/\hbar=13.5\,\mu\mathrm{m}^{-1}$, giving $|k_{\mathrm{col},0}a_{\uparrow}|=0.71$ at $a_{\uparrow}/a_0=-1000$. The finite-energy correction is therefore significant in our experiment. Higher-partial-wave contributions are neglected because the product of $k_{\mathrm{col},0}$ and the van der Waals radius is still small. We also neglect effective-range corrections for the broad Feshbach resonances considered here.

Carrying out the integration with the finite-energy cross section yields
\begin{equation}
	F_\mathrm{drag} = \sigma(k_\mathrm{col}) n_\mathrm{B} \left[ \frac{\mu^2 u^2}{m_\mathrm{B}} F_c + \frac{m_\mathrm{B}^4}{2 \mu^2 m_\mathrm{I} u^2} \mathcal{G} \right],
	\label{eq:Drag_Force_Full}
\end{equation}
where the functions $F_c(u)$ and $\mathcal{G}(u)$ are
\begin{equation}
	F_c \equiv \left( \frac{1 - \sqrt{1 - [1 - (m_\mathrm{I}/m_\mathrm{B})^2](1 - c^2/u^2)}}{1 - m_\mathrm{I}/m_\mathrm{B}} \right)^4,
	\label{eq:Fc_Definition}
\end{equation}
\begin{equation}
	\begin{split}
		\mathcal{G} \equiv{}& X(2X^2 - 3c^2)\sqrt{X^2 + c^2} \\
		&+ 3c^4 \ln\frac{X + \sqrt{X^2 + c^2}}{c},
	\end{split}
	\label{eq:G_Definition}
\end{equation}
and $X \equiv \mu u F_c^{1/4} / m_\mathrm{B}$. In the heavy-impurity limit $m_\mathrm{B}/m_\mathrm{I} \to 0$, Eq.~\eqref{eq:Drag_Force_Full} reproduces the three-dimensional result of Ref.~\cite{2004Pitaevskii} when $\sigma(k_\mathrm{col})$ is replaced by the zero-energy cross section $\sigma_0$.

\begin{figure}
	\centering
	\includegraphics{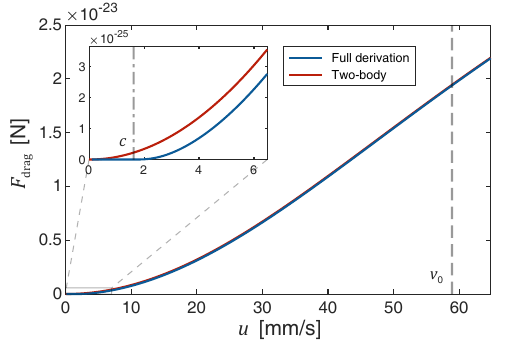}
	\caption{Drag force on an impurity moving through a homogeneous BEC as a function of the relative speed $u$. The blue and red curves show the predictions of Eqs.~\eqref{eq:Drag_Force_Full} and \eqref{eq:Drag_Force}, respectively, for $a_{\uparrow}/a_0=-1000$ and $n_\mathrm{B}=9.9\times10^{12}\,\mathrm{cm}^{-3}$. The gray dashed vertical line marks the initial relative speed $v_0$. The inset shows a close-up view near $u \sim c$, with the gray dash-dotted line marking the speed of sound of the condensate.}
	\label{FigureS_Drag_Force_Derivation}
\end{figure}

Fig.~\ref{FigureS_Drag_Force_Derivation} compares the result of Eq.~\eqref{eq:Drag_Force_Full} with the semiclassical two-body prediction of Eq.~\eqref{eq:Drag_Force} for $a_{\uparrow}/a_0=-1000$ and the mean condensate density in our experiment. The correction due to Bogoliubov suppression is pronounced near $u\sim c$ but becomes negligible in the highly supersonic regime relevant to most of our analysis. At the initial relative speed $u=v_0$, the predicted drag force is only $0.31\%$ smaller than the two-body result, justifying the use of the simpler two-body collision model in place of the more general expression under our experimental conditions.

\subsection{Effect of the force by optical dipole trap}

During the interaction time $t_1$, both the 1064\,nm ODT and the 775\,nm K-selective guide remain on [Fig.~\ref{FigureS_Sequence}]. As the Na bath accelerates the K impurities away from the trap center along $+\hat{\mathbf{x}}$, part of the work done by the drag force is converted into potential energy rather than kinetic energy. We estimate the resulting correction to the impurity velocity $v_\mathrm{I}(t_1)$ measured during $t_2$ and show that it is negligible under our experimental conditions.

We estimate this correction at a representative interaction of $a_{\uparrow}/a_0 = -1000$. For an impurity cloud whose COM is displaced by $\delta_\mathrm{COM}$ along $\hat{\mathbf{x}}$ at the end of $t_1$, the potential energy increase is $\Delta E_\mathrm{ODT} = U(\delta_\mathrm{COM} \hat{\mathbf{x}}) - U(\mathbf{0})$, where $U$ is the combined potential of the ODT and the K-selective guide. For long $t_1$, $\delta_\mathrm{COM}$ reaches several hundred $\mu$m, beyond the harmonic region of the trap. We therefore evaluate $U$ using the full Gaussian beam profiles rather than the harmonic approximation.

Adding this energy increase to the measured kinetic energy gives an estimate of the impurity velocity in the absence of the optical potential, $v_\mathrm{I, free} = \sqrt{v_\mathrm{I}^2 + 2 \Delta E_\mathrm{ODT} / m_\mathrm{I}}$. For example, at $t_1 = 1\,\mathrm{ms}$, $\Delta E_\mathrm{ODT} = k_\mathrm{B} \times 127\,\mathrm{nK}$, compared with an impurity kinetic energy of $k_\mathrm{B} \times 3.2\,\mu\mathrm{K}$, resulting in a relative velocity correction of order $10^{-2}$. Here, $k_\mathrm{B}$ is the Boltzmann constant.

At this interaction strength, the estimated relative correction to $v_\mathrm{I}$ remains below $6\%$ over the full range of $t_1$. The correction grows with the accumulated displacement and is largest at the longest $t_1$, but remains below $2\%$ for $t_1\leq1\,\mathrm{ms}$. These corrections are comparable to the statistical uncertainties of the individual $v_\mathrm{I}(t_1)$ measurements. We find similarly small corrections at all other interaction strengths. Hence, no correction for the optical trapping potentials is applied to the velocity data presented in the main text.

The corresponding cost for the Na bath is likewise negligible. The guide exerts essentially no potential on Na, while the 1064\,nm ODT confinement is weaker for Na than for K. Together with the large bath velocity $v_0$, this leaves the bath motion nearly unaffected. A separate measurement of a launched Na condensate with the 1064\,nm ODT turned on gives an average velocity of $56.7(1)\,\mathrm{mm/s}$ over the typical interaction-time window $t_1 < 5\,\mathrm{ms}$. This gradual deceleration makes the frame comoving with the bath slightly noninertial. However, the inferred average bath deceleration is more than 20 times smaller than the drag-induced impurity deceleration over the same interval, and is therefore neglected in the analysis of the main text.

\end{document}